\documentclass[12pt]{article}
\usepackage{graphicx}
\begin{document}
\rightline{NKU-03-SF3}
\bigskip
\begin{center}
{\Large\bf Scalar Perturbations of  Charged  Dilaton Black Holes}

\end{center}
\hspace{0.4cm}
\begin{center}
Sharmanthie Fernando \footnote{fernando@nku.edu} \& Keith Arnold \footnote{Arnold@nku.edu}\\
{\small\it Department of Physics \& Geology}\\
{\small\it Northern Kentucky University}\\
{\small\it Highland Heights}\\
{\small\it Kentucky 41099}\\
{\small\it U.S.A.}\\

\end{center}

\begin{center}
{\bf Abstract}
\end{center}

\hspace{0.7cm} 

We have studied the scalar perturbation of static charged dilaton black holes in 3+1 dimensions. The black hole considered here is a solution to the low-energy string theory in 3+1 dimensions. The  quasinormal modes for the scalar  perturbations are calculated using the third order WKB method. The dilaton coupling constant has a considerable effect on the values of quasi normal modes. It is also observed that there is a linear relation between the quasi normal modes and the temperature   for large black holes.



\section{Introduction}

When a black hole is perturbed by an external field, it undergo roughly three stages of response: the first stage is related to the initial wave burst coming from the source of perturbation and is independent of the properties of the black hole. The second stage is related to damped oscillations with complex frequencies and such modes are called quasi-normal modes (QNM). Frequencies of QNM are entirely dependent on the properties of the black hole such as the mass, charge and the angular momentum. The last stage of the response corresponds to a power-law decay rate at very late time which is caused by the back scattering by the gravitational field. Since QNM frequencies only depend on the physical properties of the black hole, if the radiation due to such modes are detected by gravitational antennas, it would lead to identifying the properties of black holes.
There are many works  on QNM's in various black-hole backgrounds in the literature. An excellent review is given   by Kokkotas et. al. \cite{kok1}.

QNM frequencies in black holes which are asymptotically anti-de Sitter have attracted special attention due to the well known AdS/CFT duality, which is a duality between string theory and field theory. It is conjectured that the imaginary part of the QNM frequencies which gives the time scale for the perturbation to decay is related to the time scale of the CFT on the boundary to reach thermal equilibrium. There are  many work on AdS black holes in all  dimensions on this subject \cite{chan}\cite{horo} \cite{car1} \cite{moss} \cite{wang}  \cite{kok2} \cite{kon1} \cite{kon2} \cite{kon3} \cite{kon4} \cite{kon5} \cite{li} \cite{aros}  \cite{car2} \cite{abd}.

Another reason to study QNM comes from Loop   Quantum Gravity.
It has been proposed that the asymptotic behavior of high overtones of QNM's capture important information about the quantum nature of black holes in general\cite{mot}. These proposals have lead to important observations in terms of thermodynamic properties of black holes.

In this paper we focus on a black hole arising in low energy string theory. Since string theory has become the leader in a solution to the quantum gravity issue, we would like to understand the nature of QNM's of these stringy black holes by computing them numerically.

The paper is presented as follows: In section 2 the black hole solutions are introduced. In section 3 the scalar perturbations are given. In section 4 we will computer the QNM's and discuss the results. Finally, the conclusion is given in section 5.


\section{Static Charged Dilaton Black Hole Solutions}

In this section we will give an introduction to the static charged dilaton black hole in 3+1 dimensions found by Gibbons et.al. \cite{gib} and Garfinkle et. al.\cite{gar}. The action for the Einstein-Maxwell-dilaton gravity considered is given by,

\begin{equation}
S = \int d^4x \sqrt{-g} \left[ R - 2 (\bigtriangledown \Phi )^2 -
e^{-2 a \Phi} F_{\mu \nu} F^{\mu \nu}  \right]
\end{equation}
Here $\Phi$ is the dilaton field, $R$ is the scalar curvature and $F_{\mu \nu}$ is the Maxwell's field strength. $a$ is a dimensionless parameter which we assume to be non-negative. When $a=0$, the theory gives Einstein-Maxwell gravity coupled to a free scalar. $a=1$ is the value suggested by superstring theory. The equation of motion derived from the above action in eq.(1) are,

\begin{equation}
R_{\alpha \beta} = e^{-2 a \Phi} \left( 2 F_{\alpha \gamma} F_{\beta} ^{\gamma} + \frac{1}{2} F^2 g_{\alpha \beta}\right) + 2 \partial _{\alpha} \Phi \partial _{\beta} \Phi
\end{equation}

\begin{equation}
\bigtriangledown^{\mu} \bigtriangledown_{\mu} \Phi - \frac{1}{2} a e^{-2 a \Phi} F^2 =0
\end{equation}
\begin{equation}
\bigtriangledown_{\mu} ( e^{-2a \Phi} F^{\mu \nu})=0
\end{equation}
Gibbons et.al. \cite{gib}\cite{gar} have found static charged black hole with spherical symmetry. Such black holes have the form,
\begin{equation}
ds^2 = -f(r) dt^2 + f(r)^{-1} dr^2 +  R(r)^2 ( d \theta^2 + Sin^2(\theta) d \phi)
\end{equation}
$$f(r)= \left( 1 - \frac{r_{+}}{r} \right) \left( 1 - \frac{r_{-}}{r} \right)^{\frac{1-a^2}{1+a^2}}$$

\begin{equation}
R(r)^2 = r^2 \left( 1 - \frac{r_{-}}{r} \right)^{\frac{ 2 a^2}{1 + a^2}}
\end{equation}
Here, $r_+$ and $r_-$ are the outer and inner horizons respectively. The mass $M$ and the charge $Q$ of the black hole are related to $r_+$ and $r_-$ as follows:
$$ 2 M = r_+ + \left( \frac{1-a^2}{1+a^2} \right) r_-
$$
\begin{equation}
Q^2 = \frac{ r_+ r_-} { 1 +a^2}
\end{equation}
The dilaton field and the Maxwell field strength is given by,
$$ e^{2 a \Phi} = \left( 1 - \frac{r_{-}}{r} \right) ^{\frac{ 2 a^2}{1 + a^2}}$$
\begin{equation}
F_{t r} = \frac{e^{ 2 a \Phi} }{R^2}
\end{equation}
The Hawking temperature of the black hole is given by,
\begin{equation}
T = \frac{1}{4 \pi r_+ } \left( \frac{r_+ - r_-}{r_+} \right)^{\frac{1-  a^2}{ 1 + a^2}}
\end{equation}


\section{Scalar Perturbation of Charged Black Holes}

In this section we will develop the equations for a scalar field in the background of the charged dilaton black hole introduced in the previous section. The  equation for a massless scalar field in curved space-time is  written as,
\begin{equation}
\bigtriangledown ^2 \psi  =0
\end{equation}
which is  equal to,
\begin{equation}
\frac{1}{\sqrt{-g}} \partial_{\mu} ( \sqrt{-g} g^{\mu \nu} \partial_{\nu} \psi ) =0
\end{equation}
Using the ansatz,
\begin{equation}
\psi =  e^{- i \omega t} Y(\theta,\phi) \frac{\eta(r)}{r} 
\end{equation}
eq.(11) leads to the radial equation,
\begin{equation}
\left(\frac{d^2}{dr_{*}^2} + \omega^2 \right) \eta(r) = V(r_*) \eta(r)
\end{equation}
where,
\begin{equation}
V(r) =  \frac{l(l+1)}{R^2} f + \frac{f f' R'}{R} + \frac{f^2 R''}{R}
\end{equation}
and $r_*$ is the  ``tortoise'' coordinate given by,
\begin{equation}
dr_{*} = \frac{dr}{f}
\end{equation}
Note that $l$ is the spherical harmonic index. In this paper we will mainly focus on black holes with $a=1$. In that case, $r_*$ and $r$ has the following relations,
\begin{equation}
r_* = r + r_+ ln( r - r_+)
\end{equation}
Hence when $r \rightarrow \infty$, $r_* \rightarrow \infty$ and when $r \rightarrow r_+$, $r_* \rightarrow - \infty$.


\section{Quasi normal Modes of the Charged Dilaton Black Hole}

Quasinormal modes of a black hole are obtained by solving the perturbation equation with boundary conditions imposed. The boundary conditions for asymptotically flat black holes are:  purely ingoing waves at the horizon and purely out going waves at the spatial infinity. Once these conditions are imposed, the QNM frequencies can be found which are complex.  Quasi normal modes for the above black  hole for gravitational perturbations have been studied by Ferrari et. al. \cite{fera}  and Konoplya\cite{kon5}.

In general, it is not possible to solve the perturbation equation given in eq.(13)  analytically.  There are very few cases where analytical solutions were obtained. Examples are,  BTZ black hole \cite{bir1}  and the charged dilaton black hole \cite{fer} in 2+1 dimensions. In five dimensions,  analytical  values have been obtained for vector perturbations by Nunez and Starinets \cite{nun}.

There are several approaches  to compute quasi normal modes in the literature. Here, we will follow a semi analytical technique developed by Iyer and Will \cite{will}. The method makes use of the WKB approximation  carried out to third order beyond the eikonal approximation. This approach has been applied to the Schwarzschild \cite{iyer1} Reissner-Nordstrom \cite{koko1} and for gravitational perturbations of the charged dilaton black hole \cite{ fera} \cite{kon5}. We will first review the basics of this method as follows;

Let us rewrite the perturbation eq.(13) in the following form.
\begin{equation}
\left(\frac{d^2}{dr_{*}^2} + \omega^2 - V(r_{*}) \right)  \eta(r_{*}) = 0
\end{equation}
Then, one can define new variables $ \Lambda_2, \Lambda_3$,  and $ \alpha $ as follows.
\begin{equation}
\Lambda_2 = \frac{1}{(-2 V^{(2)})^{1/2}} \left[\frac{1}{8} \left[ \frac{V^{(4)}}{V^{(2)}} \right] ( \frac{1}{4} + \alpha^2 ) -\frac{1}{288} \left[\frac{V^{(3)}}{V^{(2)} } \right]^2 ( 7 + 60 \alpha^2 ) \right]
\end{equation}
$$
\Lambda_3= \frac{ \alpha}{(-2 V^{(2)})} \left[ \frac{5}{6912} \left[ \frac{ V^{(3)}}{V^{(2)}} \right]^4 ( 77 + 188 \alpha^2) - \frac{1}{384} \left[ \frac{(V^{(3)})^2 V^{(4)} }{ (V^{(2)})^3} \right] ( 51 + 100 \alpha^2) \right.
$$
\begin{equation} 
\left. + \frac{1}{2304} \left[ \frac{V^{(4)}}{V^{(2)}} \right]^2 ( 67 + 68 \alpha^2) + \frac{1}{288} \left[ \frac{V^{(3)} V^{(5)}}{(V^{(2)})^2} \right] ( 19+ 28 \alpha^2) 
-\frac{1}{288} \left[ \frac{V^{(6)}}{ V^{(2)} } \right] 
( 5 + 4 \alpha^2)\right] 
\end{equation}

\begin{equation}
\alpha= n + \frac{1}{2}
\end{equation}
Note that the superscript $(i)$ denotes the appropriate number of derivatives of $V(r_*)$ with respect to $r_*$ evaluated at the maximum of $V(r_*)$. In the case of black hole perturbations where $V(r_*)$ is independent of frequency $\omega_n$, the quasi normal mode frequencies are given by,
\begin{equation}
i( \omega^2_n-  V) =   + \sqrt{( -2 i V^{(2)})} \left( \Lambda_2 + \Lambda_3 + \alpha \right) 
\end{equation}
We will represent $\omega = \omega_R - i \omega_I$.
We calculated  the  quasi normal modes $\omega(0)$ corresponding to the lowest mode ($n=0$) of the dilaton charged  black holes. First, we vary the value of $a$ for $r_- =1$ and $r_+=3$ to observe the behavior with respect to the dilaton coupling as follows:

\begin{center}
\begin{tabular}{|l|l|l|r} \hline \hline
 a & $\omega_R$  &  $\omega_I$ \\ \hline
0 & 0.08809 & 0.01220 \\ \hline
0.33 & 0.09350 & 0.01625 \\ \hline
.5 & 0.09900 & 0.01782 \\ \hline

1 & 0.11830 & 0.02022 \\ \hline
1.33 & 0.12899 & 0.01990 \\ \hline
2 & 0.14287 & 0.01873 \\ \hline
3 & 0.15260 & 0.01771 \\ \hline
4 & 0.15686 & 0.01725 \\ \hline
8 & 0.16153 & 0.01675 \\ \hline

\end{tabular}
\end{center}

\begin{center}

\scalebox{.9}{\includegraphics{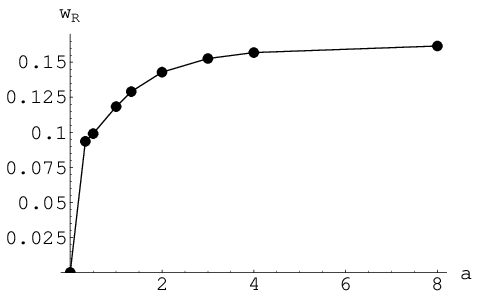}}

\vspace{0.3cm}
{Figure 1. The behavior of Re $\omega$ with the dilaton coupling $a$ for $r_-=1$ and $r_+=3$.}

\vspace{0.3 cm}

\scalebox{.9}{\includegraphics{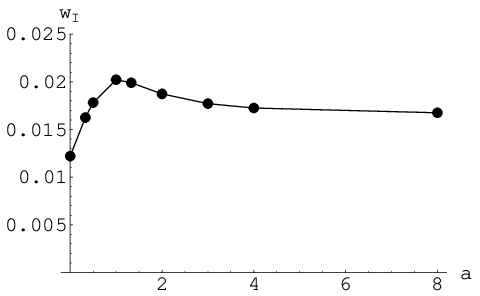}}

\vspace{0.3cm}
{Figure 2. The behavior of Im $\omega$ with the dilaton coupling $a$ for $r_-=1$ and $r_+=3$.}

\end{center}

The behavior of $ \omega $ vs.  $ a$ is given in  Figure 1 and 2. 
It is clear that Re $\omega$ increase with $a$  to a stable value. On the other hand Im $\omega$ increase up to $a =1$ and then decrease to a stable value as given in the Figure 2. It seems $a=1$ has the greatest damping in comparison with other values. Also note that oscillations are higher for larger values of $a$. There is  considerable difference in QNM's in comparison with the Reissner-Nordstrom black hole given for $a=0$. In contrast, for gravitational perturbations, not much of a difference was observed between the values for Reissner-Nordstrom and the other stringy black holes \cite{kon5}.

The behavior of these graphs for large $a$ can be given a simple explanation. When $a$ increases, the metric approaches to a black hole with $f$ and $R$ given in eq.(6)  to be
$$
f(r)= \left( 1 - \frac{r_{+}}{r} \right) \left( 1 - \frac{r_{-}}{r} \right)^{-1}
$$
\begin{equation}
R(r)^2 = r^2 \left( 1 - \frac{r_{-}}{r} \right) ^{2}
\end{equation}
Hence the effective potential $V(r)$ would be independent of $a$ for large values of $a$ leading to constant QNM's.

Next we calculate the QNM's for the spherical  harmonic  index $l=0,1,2$. We have only considered $a=1$ case of the dilaton coupling which is special from superstring point of view. The potential $V(r)$ for $l=0,1,2$ values are given in Figures 3,4,5 for $a=1$.

\vspace{0.3 cm}
\begin{center}

\scalebox{.9}{\includegraphics{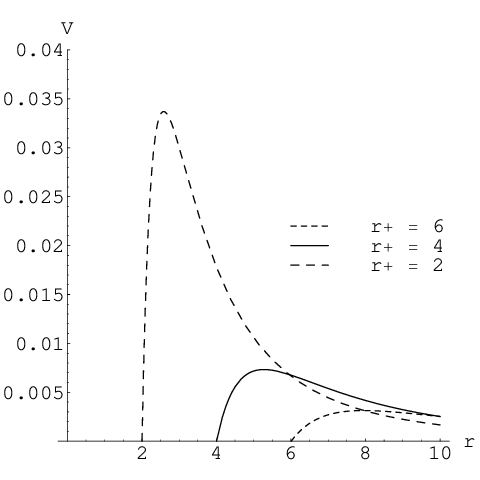}}

{Figure 3. The effective potential $V(r)$ for $l=0$. The inner horizon  $r_-$=$1$ in all three cases.}

\vspace{0.3 cm}

\scalebox{.9}{\includegraphics{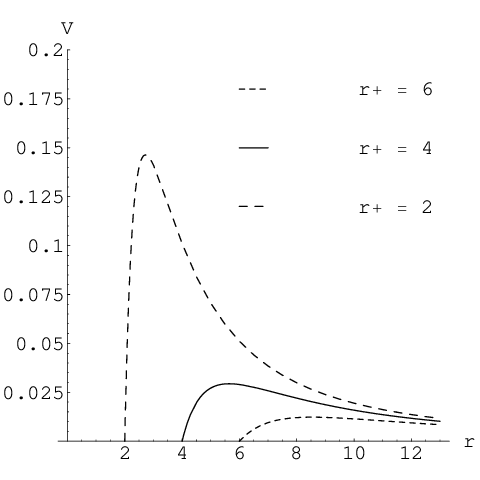}}

\vspace{0.3cm}
{Figure 4. The effective potential $V(r)$ for $l=1$. The inner horizon  $r_-$=$1$ in all three cases.}

\vspace{0.3cm}

\scalebox{.9}{\includegraphics{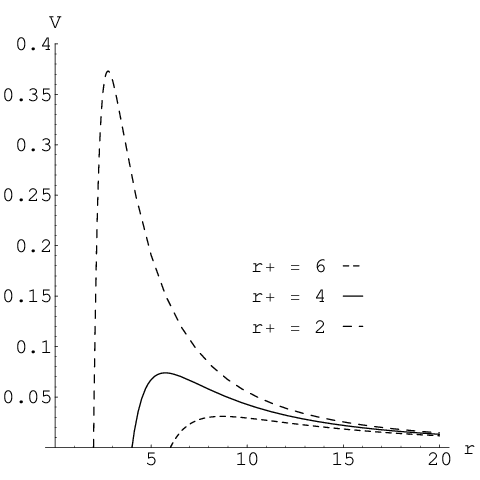}}

\vspace{0.3cm}
{Figure 5. The effective potential $V(r)$ for $l=2$. The inner horizon  $r_-$=$1$ in all three cases.}

\vspace{0.3 cm}

\end{center}
First we give the quasi normal modes for $l=0$ as follows:
\\

\begin{center}
\begin{tabular}{|l|l|l|r} \hline \hline
 $r_+$ & $\omega_R$  &  $\omega_I$ \\ \hline
2 & 0.18757 & 0.04034 \\ \hline
4 & 0.08662 & 0.01338 \\ \hline
6 & 0.05650 & 0.00760 \\ \hline

8 & 0.04195 & 0.00529 \\ \hline

10 & 0.03366 & 0.00404 \\ \hline

20 & 0.01650 & 0.00184 \\ \hline

50 & 0.00656 & 0.00069 \\ \hline

\end{tabular}
\end{center}

\begin{center}
\vspace{0.3 cm}

\scalebox{.9}{\includegraphics{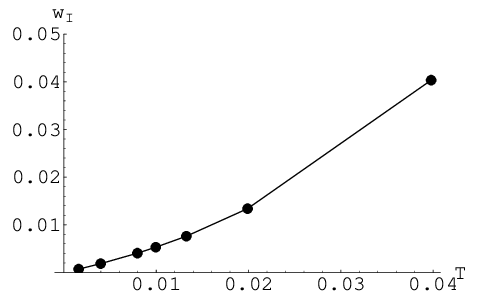}}

\vspace{0.3cm}
{Figure 6. The behavior of $\omega_I$ with the Temperature for  $l=0$. The inner horizon  $r_-$=$1$.}

\newpage

\scalebox{.9}{\includegraphics{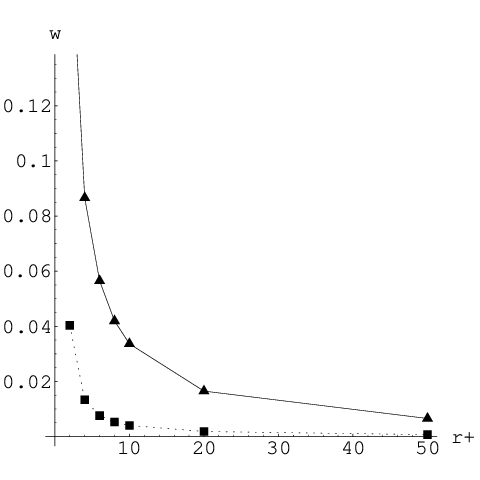}}

\vspace{0.3cm}
{Figure 7. The behavior of $\omega$ with $r_+$ for  $l=0$. The inner horizon  $r_-$=$1$. The dark lines are Re $\omega$ and dashed lines are for Im $\omega$.}

\vspace{0.3 cm}

\end{center}

The plot of $Im (\omega)$ vs temperature is given in the Figure 6.
One can see a linear behavior  of Im $\omega$ for large black holes. Note that the temperature of the  black hole for $a=1$ is $T= 1 / 4 \pi r_+ $. Hence larger the black hole smaller the temperature $T$. We have kept $r_-$ constant in these studies. We have also given the behavior of Re $\omega$ and Im $\omega$ with $r_+$ in the Figure 7. Note that the QNM's decrease with $r_+$. This behavior is somewhat different to the behavior of QNM's in gravitational perturbations given by Konoplya \cite{kon5}. There,  Im $\omega$  increase for small $Q$ and  decrease for large $Q$.

Second, we present the quasi normal modes for $l=1$ as follows:

\begin{center}
\begin{tabular}{|l|l|l|r} \hline \hline
 $r_+$ & $\omega_R$  &  $\omega_I$ \\ \hline
2 & 0.39575 & 0.10147 \\ \hline
4 & 0.17760 & 0.04655 \\ \hline
6 & 0.11491 & 0.03025 \\ \hline

8 & 0.08497 & 0.02242 \\ \hline

10 & 0.06742 & 0.01780 \\ \hline

20 & 0.03318 & 0.00878 \\ \hline

50 & 0.01315 & 0.00348 \\ \hline

\end{tabular}
\end{center}

\vspace{0.3 cm}

\begin{center}

\scalebox{.9}{\includegraphics{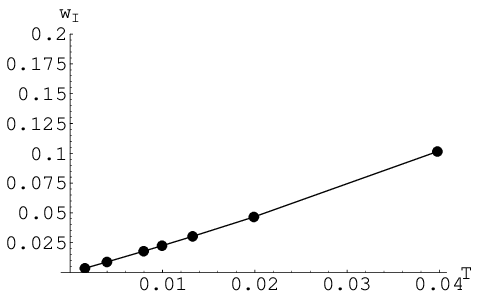}}

\vspace{0.3cm}
{Figure 8. The behavior of $\omega_I$ with the Temperature for  $l=1$. The inner horizon  $r_-$=$1$.}

\vspace{0.3 cm}

\scalebox{.9}{\includegraphics{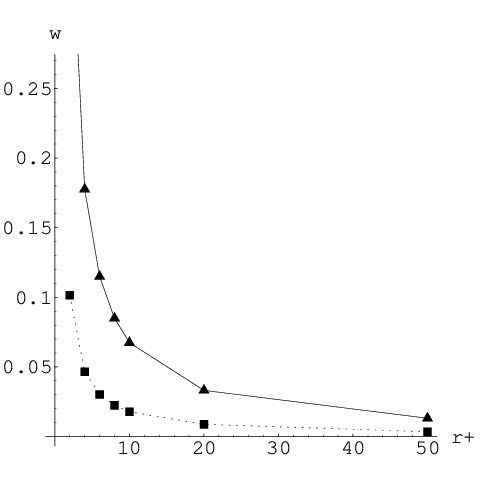}}

\vspace{0.3cm}
{Figure 9. The behavior of $\omega$ with $r_+$ for  $l=1$. The inner horizon  $r_-$=$1$. The dark lines are Re $\omega$ and dashed are for Im $\omega$.}

\vspace{0.3 cm}

\end{center}
The plot of $Im(\omega)$ vs temperature is given in the Figure 8. For this value of $l$ the  linear relation is evident. Also the behavior of $\omega$ is given in Figure 9 which has similar behavior as for $l=0$.

Last, we present the quasi normal modes for $l=2$. 
\begin{center}
\begin{tabular}{|l|l|l|r} \hline \hline
 $r_+ $ & $\omega_R$  &  $\omega_I$ \\ \hline
2 & 0.62232 & 0.11964 \\ \hline
4 & 0.27766 & 0.05645 \\ \hline
6 & 0.17936 & 0.03699 \\ \hline

8 & 0.13254 & 0.02751 \\ \hline

10 & 0.10512 & 0.16982\\ \hline

20 & 0.22809 & 0.02190 \\ \hline

50 & 0.02047 & 0.00431 \\ \hline

\end{tabular}
\end{center}
\vspace{0.3 cm}

\begin{center}

\scalebox{.9}{\includegraphics{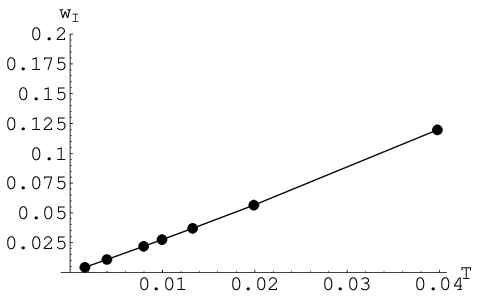}}

\vspace{0.3cm}
{Figure 10. The behavior of $\omega_I$ with the Temperature for  $l=2$. The inner horizon  $r_-$=$1$.}

\vspace{0.3 cm}

\scalebox{.9}{\includegraphics{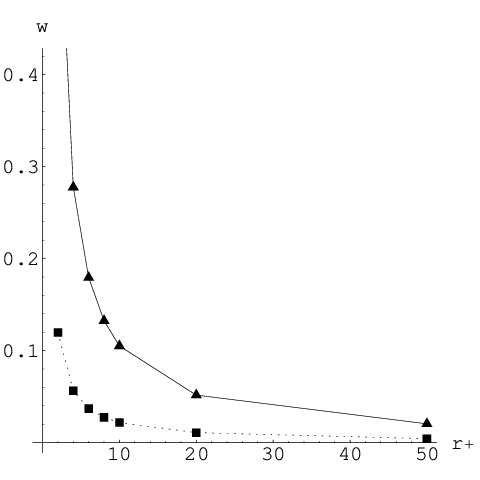}}

\vspace{0.3cm}
{Figure 11. The behavior of $\omega$ with $r_+$ for  $l=2$. The inner horizon  $r_-$=$1$. The dark lines are Re $\omega$ and dashed are for Im $\omega$.}

\vspace{0.3 cm}

\end{center}

The plot of Im ($\omega$) vs temperature is given in the Figure. 10,  where the linear relation is observed. The $\omega$ vs $r_+$ is given in Figure 11. 

For all the values of $l$ considered here, it is safe to say that there is a linear relation at least for large black holes.  This behavior is similar to the Schwarzschild anti-de-Sitter black hole studied by Horowitz and Hubeny \cite{horo}.

We have also studied the QNM's for varying $l$ with a fixed value of the horizons. The Figure. 12 and 13   gives a plot of  $\omega_R$ and $\omega_I$ with the spherical harmonic index $l$. We have chosen  $r_-=1$ and $r_+=4$ for this computation.

\begin{center}
\vspace{0.3 cm}

\scalebox{.9}{\includegraphics{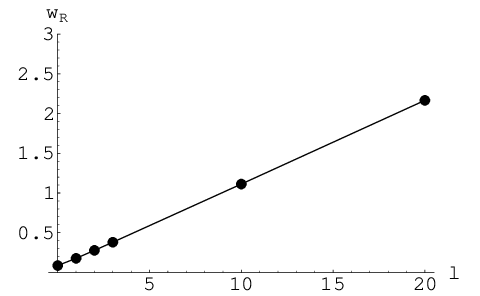}}

\vspace{0.3cm}
{Figure 12. The behavior of Re $\omega$ with the angular harmonic index $l$ for $r_+=4$ and $r_-$=$1$.}

\vspace{0.3 cm}

\scalebox{.9}{\includegraphics{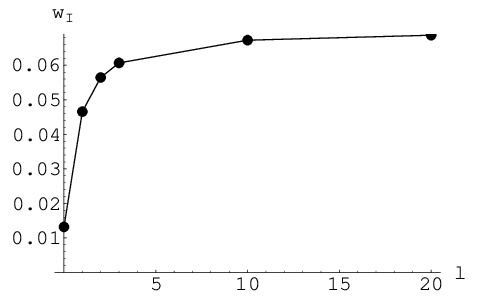}}

\vspace{0.3cm}
{Figure 13. The behavior of Im $\omega$ with the angular harmonic index $l$ for $r_+=4$ and $r_-$=$1$.}

\end{center}

Note that  the real  part of $\omega$ increases linearly  with  $l$. On the  other hand, Im $\omega$ becomes stable for large $l$. In contrast,  the QNM's of Schwarzschild-anti-de-Sitter black holes in Horowitz and Hubeny \cite{horo} had decreasing $\omega_I$ and increasing $\omega_R$.


\section{Conclusion}

We have studied  quasi normal modes for the charged dilaton black hole with scalar perturbations. $n=0$  quasi normal modes  are computed  using the third order  WKB method.  It was observed that $a=1$ gives the maximum Im $\omega$ leading to maximum damping. One of the  main result of the paper is the supporting evidence of a linear behavior between the imaginary  part of the quasi normal modes and the temperature for large black holes. This was observed first by Horowitz and Hubeny for Schwarzschild-anti-de-Sitter black holes \cite{horo}. There, they showed a linear relation between QNM's and temperature for large black holes in several dimensions. For black holes in anti-de-Sitter space, relations between QNM's and conformal field theory of the boundary are discussed in many papers \cite{bir2}. It would be interesting to give reasons  for the behavior of the QNM's observed in this paper of low energy string theory black holes. It was also noted that when the spherical index $l$ is increased, the Re $\omega$ increases leading to greater oscillations. On the other hand Im $\omega$ approaches to a fixed value for larger $l$.

There are many methods developed to compute QNM other than the WKB method used in this paper. Horowitz and Hubeny \cite{horo} employed a power series expansion method to compute QNM's of a black hole in AdS space. A semi-analytical method was developed by Weaver\cite{wea} to compute QNM frequencies of higher overtone numbers. This method was applied to Schwarzschild and Kerr black hole. It would be interesting to use these different methods to compute QNM frequencies and compare with the results obtained in this paper.

\vspace{0.5cm}

{\bf Acknowledgments}: We like to thank Don Krug and Micah Murray for helping with   computing.

\end{document}